\documentclass[3p,times,twocolumn]{elsarticle}

\usepackage{ecrc}

\volume{00}

\firstpage{1}

\journalname{Nuclear Physics B Proceedings Supplement}

\runauth{}

\jid{nuphbp}

\jnltitlelogo{Nuclear Physics B Proceedings Supplement}

\usepackage{amssymb}

\usepackage[figuresright]{rotating}

\begin{document}
\newcommand{\bb}{\ensuremath{\beta\beta}}
\newcommand{\bbonu}{\ensuremath{\beta\beta0\nu}}
\newcommand{\bbtnu}{\ensuremath{\beta\beta2\nu}}
\newcommand{\Monu}{\ensuremath{\Big|M_{0\nu}\Big|}}
\newcommand{\Mtnu}{\ensuremath{\Big|M_{2\nu}\Big|}}
\newcommand{\Gonu}{\ensuremath{G_{0\nu}(\Qbb, Z)}}
\newcommand{\Gtnu}{\ensuremath{G_{2\nu}(\Qbb, Z)}}

\newcommand{\mbb}{\ensuremath{m_{\beta\beta}}}
\newcommand{\kgy}{\ensuremath{\rm kg \cdot y}}
\newcommand{\ckky}{\ensuremath{\rm counts/(keV \cdot kg \cdot yr)}}
\newcommand{\mbba}{\ensuremath{m_{\beta\beta}^a}}
\newcommand{\mbbb}{\ensuremath{m_{\beta\beta}^b}}
\newcommand{\mbbt}{\ensuremath{m_{\beta\beta}^t}}
\newcommand{\nbb}{\ensuremath{N_{\beta\beta^{0\nu}}}}

\newcommand{\Qbb}{\ensuremath{Q_{\beta\beta}}}

\newcommand{\Tonu}{\ensuremath{T_{1/2}^{0\nu}}}

\newcommand{\Ttnu}{\ensuremath{T_{1/2}^{2\nu}}}

\newcommand{\Xe}{\ensuremath{^{136}}Xe}

\newcommand{\TwoS}{\ensuremath{^{2}S_{1/2}}}

\newcommand{\TwoP}{\ensuremath{^{2}P_{1/2}}}

\newcommand{\TwoD}{\ensuremath{^{2}D_{3/2}}}

\newcommand{\CS}{\ensuremath{^{137}}Cs}

\newcommand{\NA}{\ensuremath{^{22}}Na}

\newcommand{\Bi}{\ensuremath{^{214}}Bi}

\newcommand{\Tl}{\ensuremath{^{208}}Tl}

\newcommand{\Pb}{\ensuremath{^{208}}Pb}
\newcommand{\PBD}{\ensuremath{^{210}}Pb}

\newcommand{\Po}{\ensuremath{^{214}}Po}

\newcommand{\bru}{cts/(keV$\cdot$kg$\cdot$y)}
\newcommand{\HPXE}{\sc{HPXe}\rm}
\newcommand{\BATA}{\sc{BaTa}\rm}

\newcommand{\minitab}[2][l]{\begin{tabular}{#1}#2\end{tabular}}

\newcommand{\thedraft}{0.1.1}

\newcommand{\MO}{\ensuremath{{}^{100}{\rm Mo}}}
\newcommand{\SE}{\ensuremath{{}^{82}{\rm Se}}}
\newcommand{\ZR}{\ensuremath{{}^{96}{\rm Zr}}}
\newcommand{\KR}{\ensuremath{{}^{82}{\rm Kr}}}
\newcommand{\ND}{\ensuremath{{}^{150}{\rm Nd}}}
\newcommand{\XE}{\ensuremath{{}^{136}\rm Xe}}
\newcommand{\GE}{\ensuremath{{}^{76}\rm Ge}}
\newcommand{\GES}{\ensuremath{{}^{68}\rm Ge}}
\newcommand{\TE}{\ensuremath{{}^{128}\rm Te}}
\newcommand{\TEX}{\ensuremath{{}^{130}\rm Te}}
\newcommand{\TL}{\ensuremath{{}^{208}\rm{Tl}}}
\newcommand{\CA}{\ensuremath{{}^{48}\rm Ca}}
\newcommand{\CO}{\ensuremath{{}^{60}\rm Co}}
\newcommand{\PO}{\ensuremath{{}^{214\rm Po}}}
\newcommand{\U}{\ensuremath{{}^{235}\rm U}}
\newcommand{\CT}{\ensuremath{{}^{10}\rm C}}
\newcommand{\BE}{\ensuremath{{}^{11}\rm Be}}
\newcommand{\BO}{\ensuremath{{}^{8}\rm Be}}
\newcommand{\UDTO}{\ensuremath{{}^{238}\rm U}}
\newcommand{\CD}{\ensuremath{^{116}{\rm Cd}}}
\newcommand{\THO}{\ensuremath{{}^{232}{\rm Th}}}
\newcommand{\BI}{\ensuremath{{}^{214}}Bi}

\begin{frontmatter}



\title{The NEXT experiment}
\author{J.J. Gomez-Cadenas \corref{r1}}
\address{Instituto de F\'isica Corpuscular (IFIC), CSIC \& Universidad de Valencia}
\ead{gomez@mail.cern.ch}
\ead[url]{http://next.ific.uv.es/next/}
\cortext[r1]{On behalf of the NEXT collaboration}
 
\dochead{}

\title{The NEXT experiment}


\author{}

\address{}

\begin{abstract}
NEXT (Neutrino Experiment with a Xenon TPC) is an experiment to search neutrinoless double beta decay processes (\bbonu). 
The isotope chosen by NEXT is  \XE. The NEXT technology is based in the use of time projection chambers operating at a typical pressure of 15 bar and using electroluminescence to amplify the signal (\HPXE). The main advantages of the experimental technique are: a) excellent energy resolution; b) the ability to reconstruct the trajectory of the two electrons emitted in the decays, a unique feature of the \HPXE\ which further contributes to the suppression of backgrounds; c) scalability to large masses; and d) the possibility to reduce the background to negligible levels thanks to the barium tagging technology (\BATA).

The NEXT roadmap was designed in four stages: i) Demonstration of the \HPXE\ technology with prototypes deploying a mass of natural xenon in the range of 1 kg; ii) Characterisation of the backgrounds to the \bbonu\ signal and measurement of the \bbtnu\ signal with the NEW detector, deploying 10 kg of enriched xenon and operating at the LSC; iii) Search for \bbonu\ decays with the NEXT-100 detector, which deploys 100 kg of enriched xenon; iv) Search for \bbonu\ decays with the BEXT detector, which will deploy masses in the range of the ton and will introduce two additional handles, only possible in a \HPXE: a) A magnetic field, capable of further enhancing the topological signal of NEXT; and b) barium-tagging (a technique pioneered by the EXO experiment which is also accessible to NEXT).  

The first stage of NEXT has been successfully completed during the period 2009-2013. The prototypes NEXT-DEMO (IFIC) and NEXT-DBDM (Berkeley) were built and operated for more than two years. These apparatuses have demonstrated the main features of the technology. The experiment is currently developing its second phase. The NEW detector is being constructed during 2014 and will operate in the LSC during 2015. The NEXT-100 detector will be built and commissioned during 2016 and 2017 and will start data taking in 2018. NEXT-100 could discover \bbonu\ processes if the period of the decay is equal or less than $6 \times 10^{25}$~year. The fourth phase of the experiment (BEXT) could start in 2020. 

\end{abstract}

\begin{keyword}
Majorana neutrinos
\sep High Pressure Xenon chamber (HPXe)
\sep Double beta decay
\sep NEXT
\sep Electroluminescence
\end{keyword}
\end{frontmatter}



\section{Introduction}
Neutrinos, unlike the other fermions of the Standard Model of particle physics, could be Majorana particles, that is, indistinguishable from their antiparticles. The existence of Majorana neutrinos would have profound implications in particle physics and cosmology. 

 If neutrinos are Majorana particles, there must exist a new scale of physics (at a level inversely proportional to the neutrino masses) that characterises the underlying dynamics beyond the Standard Model. The existence of such a new scale provides the simplest explanation of why neutrino masses are so much lighter than the charged fermions. Indeed, understanding the new physics responsible for neutrino masses is one of the most important open questions in particle physics, and could have profound implications in our comprehension of the mechanism of symmetry breaking, the origin of mass and the flavour problem. 

The discovery of Majorana neutrinos would also mean that the total lepton number is not conserved, an observation that could be linked to the origin of the matter-antimatter asymmetry observed in the Universe today. This is because the new physics responsible for neutrino masses could provide a new mechanism to generate this asymmetry, via a process called leptogenesis. Although the predictions are model dependent, two essential ingredients must be confirmed experimentally for leptogenesis to occur: 1) the violation of lepton number and 2) CP violation in the lepton sector. 

The only practical way to establish experimentally whether neutrinos are their own antiparticles, and whether lepton number is not conserved, is the detection of neutrinoless double beta decay (\bbonu). This is a hypothetical, very slow nuclear transition in which a nucleus with $Z$ protons decays into a nucleus with $Z+2$ protons and the same mass number $A$, emitting two electrons that carry essentially all the energy released (\Qbb). The process can occur if and only if neutrinos are Majorana particles. 

All the recent results published on \bbonu\ searches \cite{Agostini:2013mzu,Albert:2014awa,TheKamLAND-Zen:2014lma} have reported null results and therefore a lower limit on the period of \bbonu\ processes, \Tonu. This lower limit can be translated into an upper limit on the \emph{effective Majorana mass} of the electron neutrino defined as:
\begin{equation}
\mbb = \Big| \sum_{i} U^{2}_{ei} \ m_{i} \Big| \, ,
\end{equation}
where $m_{i}$ are the neutrino mass eigenstates and $U_{ei}$ are elements of the neutrino mixing matrix. The mass \mbb\ is related to the period through the equation:

\begin{equation}
(T^{0\nu}_{1/2})^{-1} = G^{0\nu} \ \big|M^{0\nu}\big|^{2} \ \mbb^{2} \, .
\label{eq:Tonu}
\end{equation}

In Eq.~\ref{eq:Tonu}, $G^{0\nu}$ is an exactly-calculable phase-space integral for the emission of two electrons and $M^{0\nu}$ is the nuclear matrix element (NME) of the transition, which has to be evaluated theoretically. The uncertainty in the NME affects the value of \mbb\ which can be obtained from \Tonu.

\section{The NEXT experiment and its innovative concepts}
\begin{figure}
\centering
\includegraphics[width=0.6\textwidth]{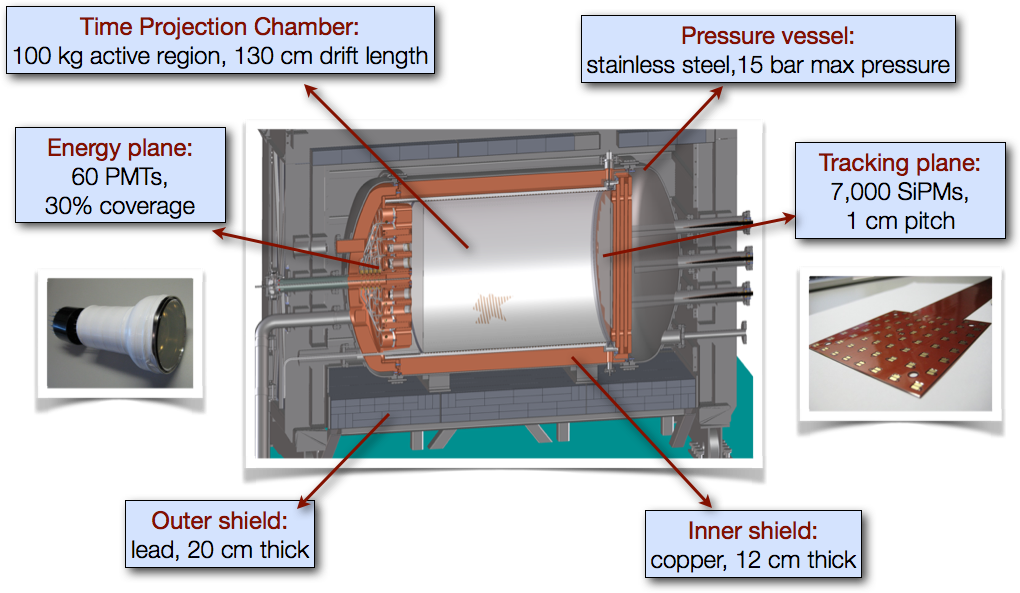}
\caption{\small A drawing of the NEXT-100 detector showing its main parts. The pressure vessel (PV) is made of a radio pure steel-titanium alloy. The PV dimensions are 130~cm inner diameter, 222~cm length, 1~cm thick walls, fot a total mass of 1\,200 kg. The inner copper shield (ICS) is made of ultra-pure copper bars and is 12~cm thick, with a total mass of 9\,000 kg. The time projection chamber includes the field cage, cathode, EL grids and HV penetrators.
The light tube is made of thin teflon sheets coated with TPB (a wavelength shifter). 
The energy plane is made of 60 PMTs housed in copper enclosures (cans).
The tracking plane is made of MPPCs arranged into dice boards (DB). 
} \label{fig.NEXT100}
\end{figure}

The \emph{Neutrino Experiment with a Xenon TPC} (NEXT) will search for \bbonu\ in \XE\ using  high-pressure xenon gas  time projection chambers 
(\HPXE)\cite{Nygren:2009zz,Granena:2009it,Alvarez:2012haa}, yielding: 
a) {\bf excellent energy resolution}, with an intrinsic limit of about 0.3\% FWHM at \Qbb, and close to that of \GE\ detectors and a demonstrated result in the vicinity of 0.5\% FWHM; b)
{\bf tracking capabilities} that provide a powerful topological signature to discriminate between signal (two electron tracks with a common vertex) and background (mostly, single electrons); c)
{\bf a fully active and homogeneous detector}, with no dead regions; d) {\bf scalability} of the technique to large masses; e) the possibility of exciting the barium ion produced in the xenon decay from the fundamental state \TwoS\ to the state \TwoP, using a ``blue'' laser (493.54 nm), and observing the ``red light'' emitted in the transition from \TwoP\ to \TwoD, thus ``tagging'' the presence of a barium atom in the xenon gas, which cannot be produced by any known background. 

The design of the NEXT-100 detector (Figure~\ref{fig.NEXT100}) is optimised for energy resolution by using proportional electroluminescent (EL) amplification of the ionisation signal\footnote{As proposed in \cite{Nygren:2009zz}}. The detection process involves the use of the prompt scintillation light from the gas as start-of-event time, and the drift of the ionisation charge to the anode by means of an electric field ($\sim0.3$ kV/cm at 15 bar) where secondary EL scintillation is produced in the region defined by two highly transparent meshes, between which there is a field of $\sim20$ kV/cm at 15 bar. The detection of EL light provides an energy measurement using photomultipliers (PMTs) located behind the cathode (the \emph{energy plane}) as well as tracking, a few mm away from production at the anode, via a dense array of silicon photomultipliers (the \emph{tracking plane}).

\subsection{NEXT prototypes}

\begin{figure}
\centering
\includegraphics[width=0.4\textwidth]{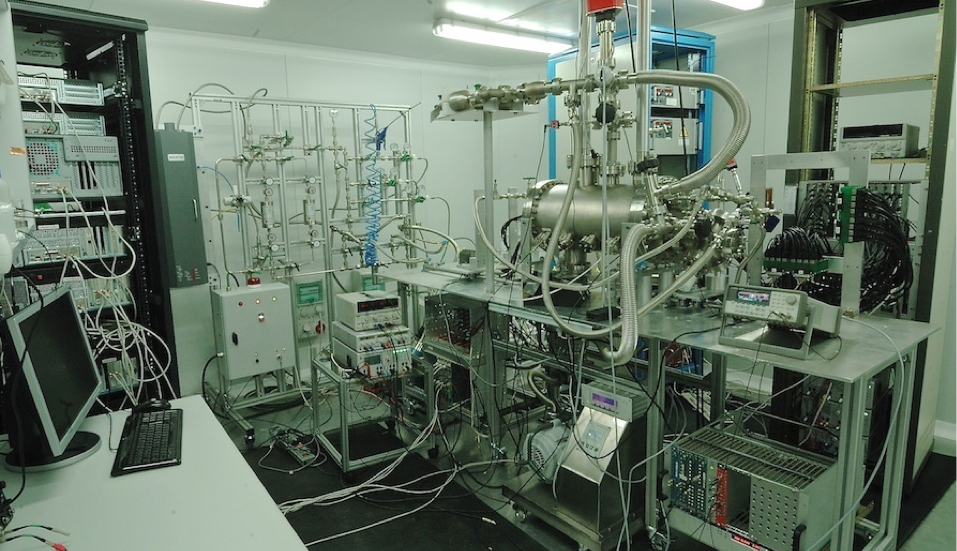}
\caption{\small The NEXT-DEMO setup at IFIC (Valencia, Spain).} \label{fig.DEMO}
\end{figure}
%

NEXT-DEMO, shown in Figure~\ref{fig.DEMO}, is as a large-scale prototype of NEXT-100. The pressure vessel has a length of 60 cm and a diameter of 30 cm. The vessel can withstand a pressure of up to 15 bar and hosts typically 1-2 kg of xenon. NEXT-DEMO is  equipped with an energy plane made of 19 Hamamatsu R7378A PMTs and a tracking plane made of 256 Hamamatsu SiPMs. 

The detector has been operating successfully for more than two years and has demonstrated: (a) very good operational stability, with no leaks and very few sparks; (b) good energy resolution ; (c) track reconstruction with PMTs and with SiPMs coated with TPB; (d) excellent electron drift lifetime, of the order of 20 ms. Its construction, commissioning and operation has been instrumental in the development of the required knowledge to design and build the NEXT detector.

The NEXT-DBDM prototype is a smaller chamber, with only 8 cm drift, but an aspect ratio (ratio diameter to length) similar to that of NEXT-100. The device has been used to perform detailed energy resolution studies, as well as studies to characterise neutrons in an \HPXE. NEXT-DBDM achieves a resolution of 1\% FWHM at 660 keV and 15 bar, which extrapolates to 0.5\% at \Qbb.

\subsection{Topological signature}

\begin{figure}
\centering
\includegraphics[width=0.5\textwidth]{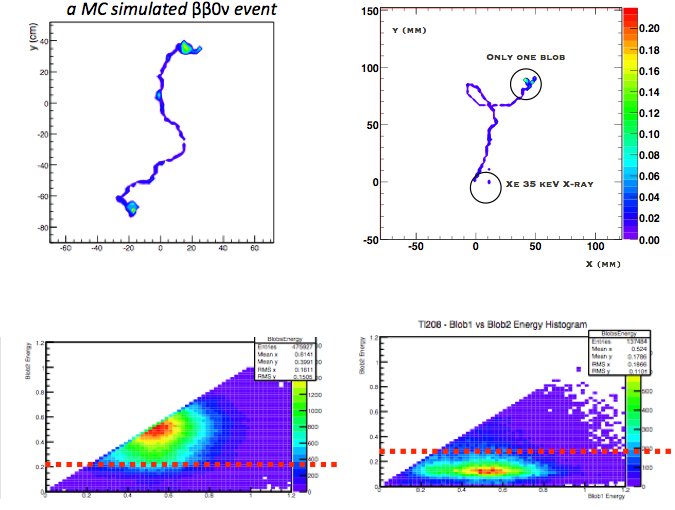}
\caption{\small NEXT has a topological signature, not available in most \bbonu\ detectors. The panel shows the reconstruction of a Monte Carlo signal (topleft) and background (bottomleft) event. The signal has two electrons (two blobs). The background has only one electron (one blob) and the associated emission of a 35 keV X-ray. The color codes energy deposition in the TPC. An scatter plot of the energy of the two blobs shows a clear separation between signal and background regions.}\label{fig.ETRK2}
\end{figure}

Double beta decay events leave a distinctive topological signature in HPXe: a continuous track with larger energy depositions (\emph{blobs}) at both ends due to the Bragg-like peaks in the d$E$/d$x$ of the stopping electrons (Figure~\ref{fig.ETRK2}, topleft). In contrast, background electrons are produced by Compton or photoelectric interactions, and are characterised by a single blob and, often, by a satellite cluster corresponding to the emission of $\sim30$-keV fluorescence x-rays by xenon (Figure~\ref{fig.ETRK2}, bottomleft).
Reconstruction of this topology using the tracking plane provides a powerful means of background rejection, as can be observed in the figure. In our TDR we chose a conservative cut to separate double--blob from single--blob events which provided a suppression factor of 20 for the background while keeping 80\% of the signal.  DEMO has reconstructed single electrons from \NA\ and \CS\ sources, as well as double electrons from the double escape peak of \TL\, demonstrating the robustness of the topological signal.

\subsection{Energy resolution}

\begin{figure}
\centering
\includegraphics[width=0.5\textwidth]{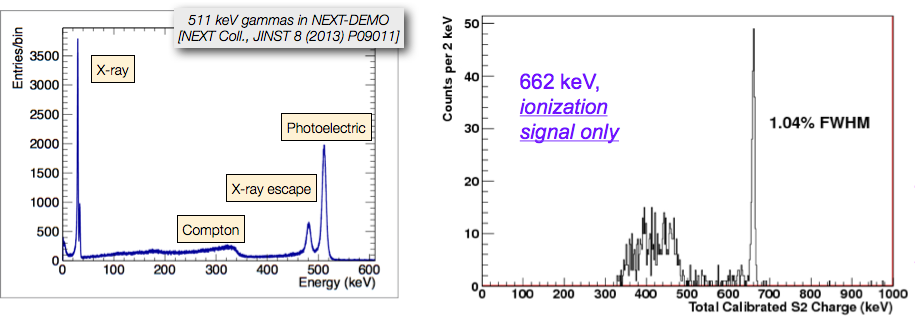}
\caption{\small Left: the full energy spectrum measured for electrons of 511 keV in the DEMO detector. Right: the spectrum near the photoelectric peak for 662 keV electrons in NEXT-DBDM. The resolution at 662 keV is 1\% FWHM (0.5\% FWHM at \Qbb). The resolution extrapolated from 511 keV is 0.7\%.}\label{fig.ERES}. 
\end{figure}

Figure~\ref{fig.ERES} shows the resolution obtained with the two NEXT prototypes. A resolution of 1\% FWHM with 
662 keV photons has been measured in the NEXT-DBDM apparatus, which extrapolates to 0.5\% FWHM at \Qbb. This result is not far from the expected limit obtained adding in quadrature the different factors that contribute to the resolution (Fano factor, photoelectron statistics and electronic noise). The resolution measured in NEXT-DEMO extrapolates to 0.7\% FWHM. The difference between both prototypes is due to better photoelectron statistics and aspect ratio in DBDM. The results, are, in any case, better than the target of 1\% FWHM described in the TDR.

The status of the NEXT experiment and the results achieved by the prototypes have been described in a recent
paper \cite{Gomez-Cadenas:2013lta}.

\subsection{\label{sec.new}The NEW detector.}

\begin{figure}
\centering
\includegraphics[width=0.5\textwidth]{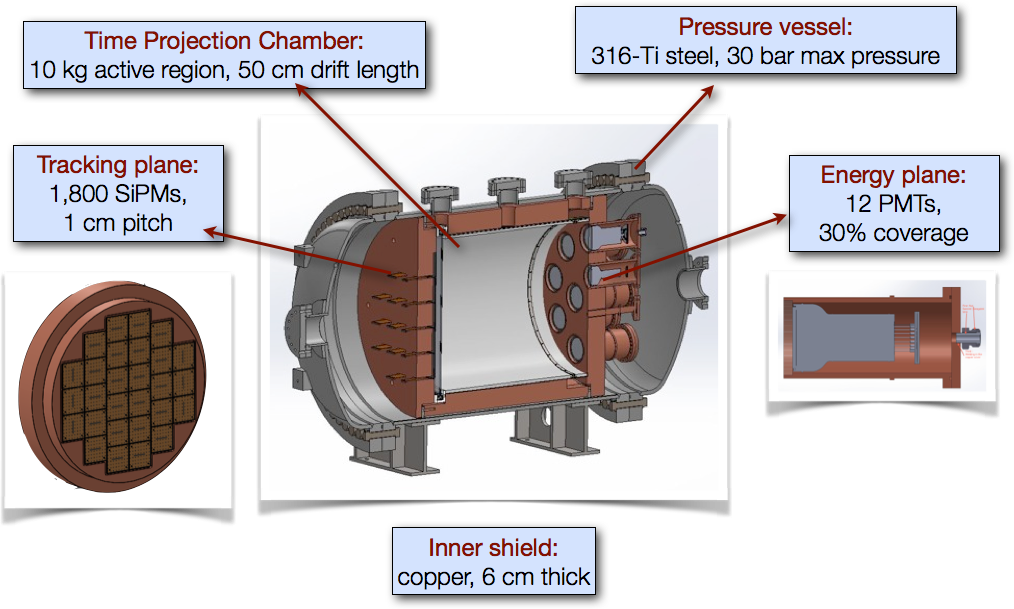}
\caption{\small The NEW apparatus.} \label{fig:NEW}
\end{figure} 

The NEW (NEXT-WHITE) apparatus\footnote{The name honours the memory of the late Professor James White, one of the key scientists of the NEXT Collaboration.}, shown in Figure~\ref{fig:NEW}, is the first phase of the NEXT detector to operate underground. NEW 
%
%
is a scale 1:2 in size (1:8 in mass) of NEXT-100. The energy plane contains 12 PMTs (20 \% of the 60 PMTs deployed in NEXT-100). The tracking plane technology consists of 30 Kapton Dice Boards (KDB) deploying 1800 SiPMs (also 20\% of the sensors). The field cage has a diameter of 50~cm and a length of 60~cm (the dimensions of the NEXT-100 field cage are roughly 1~m long and 1.2~m diameter). 

NEW is a necessary step\footnote{As formally stated by the scientific committee of the LSC, who recommended its construction in 2013.} towards the construction of NEXT-100. It will validate the technological solutions adopted by the collaboration and, as discussed below, it is essential in the definition of the project methodology. Furthermore, the NEXT background model is currently based on a sophisticated Monte Carlo simulation of all expected background sources in each part of the detector. NEW will allow the validation of the background model with actual data. 

Furthermore, the calibration of NEW with 
sources of higher energy will allow a precise study of the evolution of the resolution with the energy. 
In particular it will be plausible to measure the resolution near \Qbb\ using a thorium source, which provides 2.6 MeV gammas. Last, but not least, we intend to 
reconstruct the spectrum of \bbtnu. Those events are topologically identical to signal events (\bbonu) and can be used to demonstrate with data the power of the topological signature. 
\section{NEXT background model and expected sensitivity}

The NEXT background model describes the sources of radioactive contaminants in the detector and their activity. It allows us, via detailed simulation, to predict the background events that can be misidentified as signal and consequently, to predict the expected
sensitivity of the apparatus. A major goal of NEW is to confirm these predictions from the data themselves.

The major sources of background can be found in the 
radioactive contaminants in detector materials, in particular \BI\ and \TL\ isotopes.

After the decay of \BI, the daughter isotope, \Po, emits a number of de-excitation gammas with energies above 2.3 MeV. The gamma line at 2447 keV, of intensity 1.57\%, is very close to the $Q$-value of \XE. The gamma lines above \Qbb\ have low intensity and their contribution is negligible. 

The daughter of \TL, \Pb, emits a de-excitation photon of 2614 keV with a 100\% intensity. The Compton edge of this gamma is at 2382 keV, well below \Qbb. However, the scattered gamma can interact and produce other electron tracks close enough to the initial Compton electron so they are reconstructed as a single object falling in the energy region of interest (ROI). Photoelectric electrons are produced above the ROI but can loose energy via bremsstrahlung and populate the window, in case the emitted photons escape out of the detector. Pair-creation events are not able to produce single-track events in the ROI. 

Radon constitutes also dangerous source of background due to the radioactive isotopes $^{222}$Rn (half-life of 3.8\,d) from the $^{238}$U chain and $^{220}$Rn (half-life of 55\,s) from the $^{232}$Th chain. As a gas, it diffuses into the air and can enter the detector. \BI\ is a decay product of $^{222}$Rn, and \TL\ a decay product of $^{220}$Rn. In both cases, radon undergoes an alpha decay into polonium, producing a positively charged ion which is drifted towards the cathode by the electric field of the TPC.  As a consequence, $^{214}$Bi and $^{208}$Tl contaminations can be assumed to be deposited on the cathode surface. Radon may be eliminated from the TPC gas mixture by recirculation through appropriate filters. There are also ways to suppress radon in the volume defined by the shielding. Radon control is a major task for a \bbonu\ experiment, and will be of uppermost importance for NEXT-100. A major goal of NEW is to assess (and eventually improve) the effectiveness of radon control techniques. 

A detailed discussion of the NEXT background model has been submitted to this proceedings\footnote{M. Nebot-Guinot et al, {\bf Backgrounds and sensitivity of the NEXT double beta decay experiment.}}. 
The excellent resolution of NEXT (0.5-0.7 \% FWHM), and the combination of a low radioactive budget with a topological signature (which yields an expected background rate of $5 \times 10^{-4} \ckky$), will allow the NEXT-100 detector to reach a sensitivity to the \bbonu\ period of $\Tonu > 7 \times 10^{25}$~yr for an exposure of 300 kg$\cdot$yr. This translates into a \mbb\ sensitivity range as low as $[67-187]$~meV, depending on the NME.

\section{Towards a ton-scale high-pressure xenon TPC.}

\begin{figure}
\centering
\includegraphics[width=0.4\textwidth]{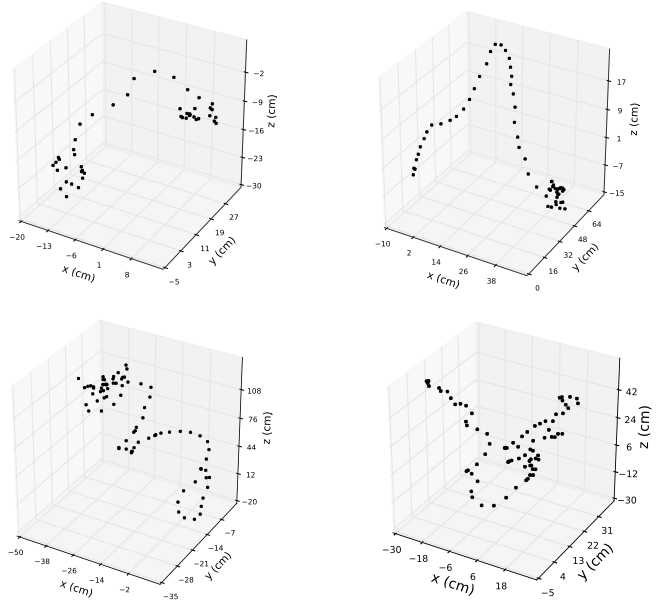}
\caption{\small Top Left: two electrons emitted in a \bbonu\ decay in the absence of magnetic field. Top right: a single (background) electron produced by the photoelectric interaction of a 2.45 MeV photon emitted in \BI\ decays, interacting in the chamber. Bottom left: two  electrons emitted in a \bbonu\ decay, turning into a double helix that originate in a common vertex, in a magnetic field of 0.5 T. Bottom right: a single background electron turning into a single helix in a magnetic field of 0.5 T. } \label{fig.KF}
\end{figure}

If no discovery is made by the current generation of experiments, the full exploration of the ``cosmologically relevant region'' (CRR), corresponding to the inverted hierarchy of neutrino masses, and \mbb\ values as low as 15~meV, requires detectors of larger mass (at least 1 ton), good resolution and extremely low specific background. The \HPXE\ technology has the potential to provide the most sensitive detector at this scale, by scaling the detector to a mass in the range of one ton and adding additional handles to further suppress the background.

\subsection*{Adding a magnetic field to enhance the topological signal}

The simplest way to improve the \HPXE\ technology is to operate the detector in a moderate magnetic field. 
In the presence of a uniform external magnetic field $B$, charged particles spiral around the field lines in circular motion with radius $r = p_T/eB$, where $p_T$~ is the momentum of the electron transverse to the direction of the field and $e$~ is the electron charge. A single energetic electron should produce a clear single spiral with radius indicative of its momentum, and a double-electron track with the same energy will produce two spirals each with much less momentum and originating from a common vertex. This information provides an additional way of separating single-electrons arising from background processes from double electrons produced in \bbonu\ decays, {\bf in spite of the large multiple scattering that the electrons suffer in a dense \HPXE}. 

This is illustrated in Figure~\ref{fig.KF}. The top-left panel shows two electrons emitted in a \bbonu\ decay in the absence of magnetic field. Notice that the vertex where both electrons originated cannot be easily measured due to multiple scattering. However, the presence of two electrons is revealed by the two blobs at the end of the tracks (regions with higher density of hits and energy deposition, which originate when each one of the electron ranges out). Instead, the top-right panel shows a single background electron, which originates inside the detector when a photon, emitted by the decay of the isotope \BI\ enters the chamber and suffers a photoelectric interaction. Notice that the single electron only displays one blob in one of the vertexes. 

However, when a sufficiently intense magnetic field is added, the separation between single and double electrons is strongly enhanced. The bottom-left panel shows again two electrons emitted in a \bbonu\ decay, turning in a magnetic field of 0.5 Tesla. The trajectory shows clearly two helices, originating from a common vertex, which becomes now visible. The bottom-right panel shows a single background electron turning in a 0.5 T magnetic field, following a single helix. Fitting the trajectories of the electrons, one can achieve an additional order of magnitude in background reduction with respect to the case without magnetic field, combining the fact that the signal is now characterised by two blobs and two helices pointing towards a common vertex.    
 
 The current (estimated) rejection factor for NEXT-100 is 0.5 counts per ton and keV in a year. If a resolution of \Qbb\ of 0.5 \% FWHM is confirmed in the large detector as we expect, this translates into 5 counts per ton in the ROI. The addition of a magnetic field may yield a value of 0.5-1 counts per ton in the ROI, thus allowing the HPXe technology to operate in the ton regime without being background limited. 

\subsection{Barium Tagging}

\begin{figure}
\centering
\includegraphics[width=0.40\textwidth]{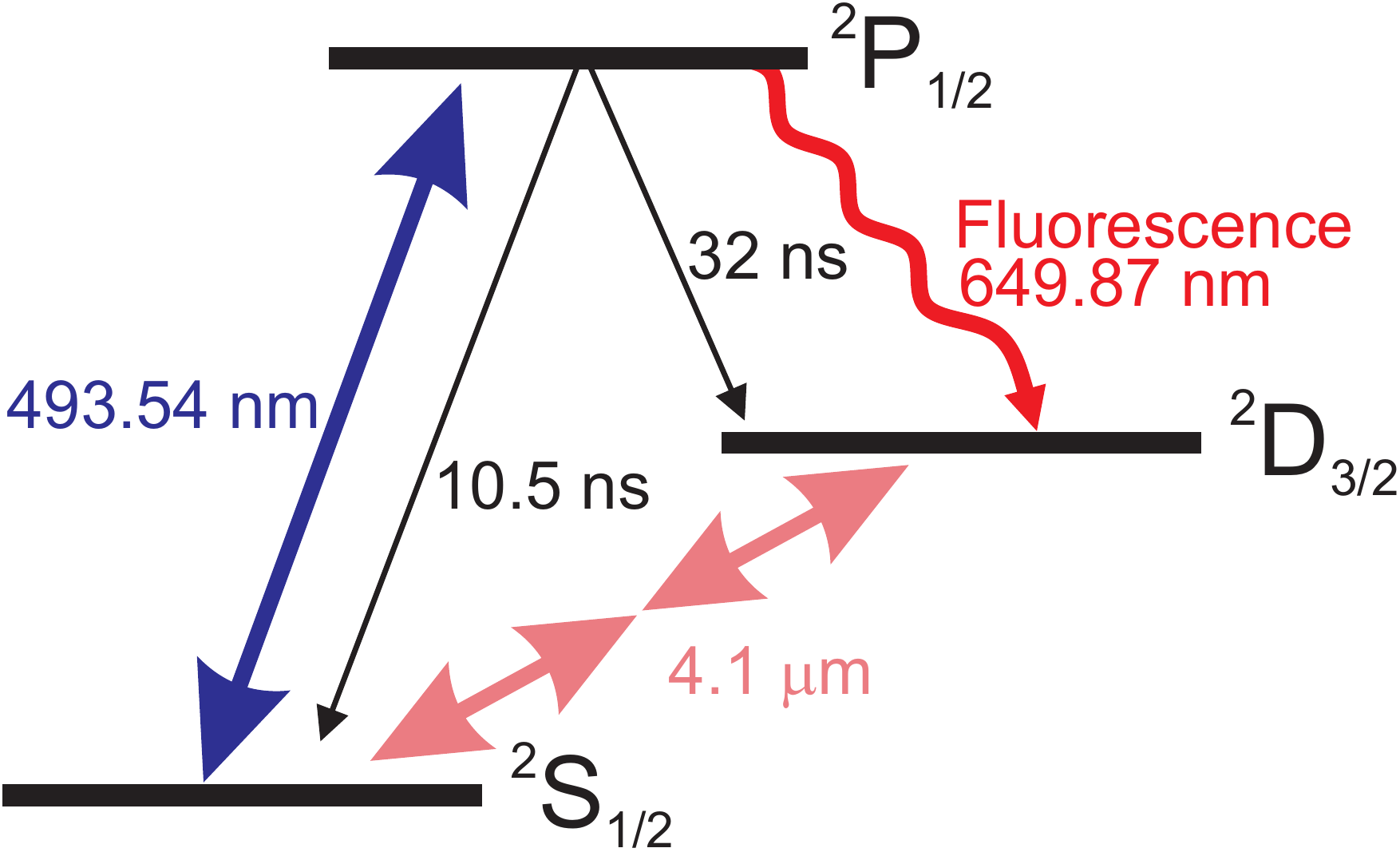}
\caption{\small The \BATA\ concept.} \label{fig.BATA}
\end{figure}

As originally suggested by Moe\footnote{M.K. Moe, ``New approach to the detection of neutrinoless double-beta decay'', Physical Review C Rapid Communications 44 (1991) R931.}, a promising possibility to further reduce the background is to develop the HPXe technology to unambiguously tag the barium ion produced in the xenon decay, $Xe \rightarrow Ba^{++} + 2 e^-$. The conceptual idea to tag $Ba^{+}$ is illustrated in Figure~\ref{fig.BATA}. A ``blue'' laser of wavelength 493.54 nm excites (``pumps'') the S state, inducing $S \rightarrow P$~transitions, with a lifetime of $\sim$ 10 ns. About 30 \% of the times the \TwoP\ states decay to the state \TwoD, emitting ``red'' (649.76 nm) fluorescence in a characteristic time of 30 ns. The state \TwoD\ is metastable, but a second laser of suitable wavelength (4.1 $\mu$m) can be used to induce the transition to the ground state (this is known as ``deshelving'').  The whole cycle takes less than 50 ns, and therefore several millions of red fluorescence photons can be emitted by a single ion. 

The practical application of this conceptual idea is by no means easy, and in fact, it has been shown to be extremely difficult in liquid xenon by the work of the EXO collaboration\cite{Dolinski:2012dta}. However, it may be feasible in a \HPXE\ detector, where a number of fortunate conditions may occur. These conditions are: a) charge reduction of the emitted barium ion, from $Ba^{++}$~to $Ba^{+}$, which can be induced by collisions with xenon atoms, or by the addition of a suitable quencher; b) ``trapping'' of the barium ion ``in situ'' by the surrounding Xe atoms, which result in a very low drift velocity for the ion; c) location of the ion, via the reconstruction of the event topology. 

All the above needs to be demonstrated with a systematic R\&D program, which must also address additional experimental issues such as pressure broadening of the laser, filtering of Rayleigh scattering, and others. The EXO collaboration has carried out extensive research of the potential for Barium Tagging, not only in a LXe TPC, but also in an \HPXE\cite{Sinclair:2011zz}. NEXT, on the other hand, has started a collaboration with CLPU\footnote{http://www.clpu.es}, a spanish national facility dedicated to ultra-intense lasers, with the aim of carrying out a systematic R\&D program to understand the potential of Barium Tagging in a high pressure gas xenon TPC. Such a program
involves a set of proof-of-concept experiments, including:

\begin{enumerate}
	\item \textbf{Ba ions generation, phase 1}. Proof-of-principle experiment with Ba ions generated by means of an electrical discharge and/or laser ablation.
	
	\item \textbf{Ba ions generation, phase 2}. Proof-of-principle experiment with Ba ions generated by an ion source.	
		
	\item \textbf{D state deshelving}. A likely scenario is that the collisional induced decay between the metastable state D and the ground state S is either not effective or too slow for obtaining an appreciable fluorescence signal. In this situation the population is trapped in the metastable state D and the fluorescence cycle can not be closed. To avoid this difficulty, deshelving the D state may be needed. A proof-of-principle experiment with an additional laser for deshelving the D state will be performed. The laser needed must have a wavelength of around 4.1\,$\mu$m. The alternative, using a red laser, would smear the characteristic red fluorescence with scattered photons from the deshelving laser.
\end{enumerate}

\subsubsection*{Proof of principle experiments}

\begin{figure}
\begin{center}
\includegraphics[width=0.5\textwidth]{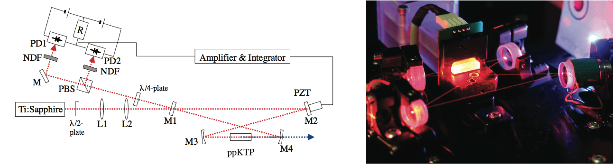}
\caption{\small Experimental set-up for resonant frequency doubling of a 
Ti:Sapphire laser using ppKTP. }
\label{fig:laser}
\end{center}
\end{figure}

In a first round of experiments we will excite resonantly the S$\leftrightarrow$P transition of Ba$^+$ ions generated by an electrical discharge between two barium electrodes and will collect the fluorescence signal of the P$\rightarrow$D transition (alternatively, laser ablation can be used). Although this generation method is not ideal because several different species other than Ba ions will be generated (e.g., BaO molecules or clusters), it does not need a major technological development.

\begin{figure}
\begin{center}
\includegraphics[width=0.5\textwidth]{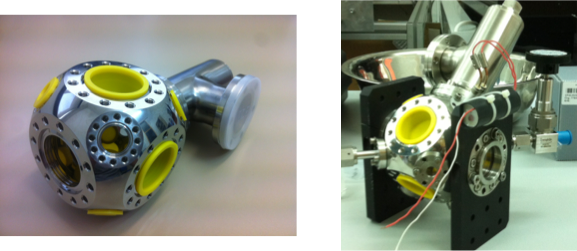}
\caption{\small The chamber for the proof-of-concept experiments, built at IFIC, and ready to be installed at CLPU.}
\label{fig:chamber}
\end{center}
\end{figure}

The laser source needed to resonantly excite the $Ba^{+}$ ions must have a wavelength of 493.5\,nm, not available in commercial lasers. The laser source will therefore be produced at the CLPU, optimising a tunable Ti:sapphire laser at 987\,nm to obtain a second harmonic generation (SHG) at 493.5\,nm, see Figure~\ref{fig:laser}. This setup allows the tuning of the wavelength and the control of the bandwidth of the laser, a necessary feature to precisely tune it to the transition frequency (e.g. to correct for pressure broadening and other effects). Figure~\ref{fig:chamber}  shows the test chamber needed for the experiments. 
We expect that this initial set of experiments will provide valuable information about the population dynamics in Ba$^+$ ions, and the influence of the different homogeneous and inhomogeneous broadening mechanisms. 

Next, we intend to generate Ba ions by an ion source that will be designed and constructed at the CLPU. This ion source will be based on selective ionisation and mass spectrometry techniques, and it will allow an efficient selection of the desired target species (e.g, $Ba^{+}$~and $Ba^{++}$). With this setup we will be able to study the recombination process Ba$^{++}\rightarrow$Ba$^{+}$ and decide whether it can be induced by collisions with xenon atoms, or whether it requires an additive. Depending on the results of the experiment, a magnetic trap can be added to improve the experimental conditions. 

Finally, we intend to perform a proof of principle experiment with an additional laser for deshelving the D state. Our approach will be to use a second laser to induce a two photon transition (one photon is forbidden by selection rules, between the states D and S, see Figure~\ref{fig.BATA}). 	

In our program we intend to reproduce and extend the pioneer work of Sinclair and others\cite{Sinclair:2011zz}, in particular focusing in the scenario of barium-tagging ``in situ'' (the approach of the EXO R\&D appears to be to extract the barium ion, both in the case of liquid and gas detectors). In any case, we believe that the barium tagging program offers an opportunity of joint development with the EXO collaboration.

Clearly the construction of a ton-scale \HPXE\ detector implementing the full Barium Tagging technology is a very challenging enterprise. On the other hand, it appears to be a promising path towards the future.

\subsection{NEXT will be BEXT}
The addition of a B-field (first stage of the upgrade) and Barium Tagging (second stage) are clear forward paths to improve the NEXT detector towards the future BEXT\footnote{B-field Experiment with a Xenon TPC, also Barium-Taggin Experiment with a Xenon TPC} apparatus, displaying  
a mass in the range of several tons. With a resolution of 0.5 \% FWHM and a background rate in the range of $10^{-4}$ (this appears relatively easy to achieve with a magnetic field) to $10^{-6}$ (as ultimately possible with barium-tagging) \ckky, it would be able to fully cover the CRR (and inverted hierarchy) region even in the most pessimistic NME scenario.

\section*{Acknowledgments}
The author would like to thank the NEXT collaboration and the CLPU group for the hard work reported in this paper. NEXT has the support of the Ministry of Economy (MINECO) under grants CONSOLIDER-Ingenio 2010 CSD2008-0037 (CUP), FPA2009-13698-C04 and FIS2012-37947-C04. NEXT has also the support of the European Research Council through the Advanced Grant 339787-NEXT.




\nocite{*}
\bibliographystyle{elsarticle-num}
\bibliography{biblio2}







\end{document}